# Use of a Virtual Laboratory to plan, execute and analyse Neutron Strain Scanning experiments.

J.A. James[1]. J.R. Santistiban[1], M.R. Daymond[2] and L. Edwards[1]

[1]Department of Materials Engineering, Faculty of Technology, The Open University, Milton Keynes MK7 6AA, UK
[2] ISIS Facility, Rutherford Appleton Laboratory, Chilton, Oxon. OX11 0QX, UK

**Abstract**

The new generation of dedicated Engineering Strain Scanners at neutron facilities such as ENGIN-X at ISIS and SMARTS at LANSCE offer considerable increases in both the throughput of samples and the density of measurements which are feasible within each sample. This trend is set to increase further with new neutron sources such as the SNS. In order to make full use of these advances the routine processes associated with setting up measurements, and analysing data need to be made as efficient as possible.

This issue has been addressed on ENGIN-X by writing a new piece of software which provides support for many of these operations. The approach is based on a virtual lab consisting of three dimensional models of the sample and lab equipment such as collimators and positioner. A typical session using the package would be; 1) Generate the sample model using primitives or from surface points measured with a coordinate measurement machine, 2) Specify fiducial and measurement points on screen, 3) Locate the sample model within the virtual and real laboratories, 4) Execute the measurement sequence using automatically generated machine control scripts, 5). Analyse the data, 6). Display data using a variety of options including superimposed on the sample model.

The inclusion of an accurate sample model within the virtual lab allows many other useful properties such as neutron path lengths and measurement gauge volumes to be determined; it is also a relatively simple matter to check for possible collisions between sample and lab equipment such as collimators thereby avoiding potentially costly mistakes.

The software which is shortly to be installed at ENGIN-X has been designed with visiting industrial and academic researchers in mind; users who need to be able to control the instrument after only a short period of training.

## 1. Introduction

This paper describes software designed to support the planning, execution and analysis of neutron strain scanning experiments on the new generation of dedicated strain scanning machines.

Stress measurement by neutron diffraction is achieved by measuring the lattice spacing within a small volume of the material known as the gauge volume. The strain within this volume is then obtained by comparing this measurement with the lattice plane spacing from an unstressed sample of the same material. The great advantage of neutron scanning methods over other strategies is that the penetration of the neutrons is such that measurements may be made deep within components and samples.

For the past decade ISIS in the UK has housed one of the worlds first dedicated neutron stress diffractometers, ENGIN, (Johnson et al (1)). As neutron stress measurement has become an increasingly important tool in engineering, pressure for beam time has increased leading to increasing over subscription rates. A successor to ENGIN, ENGIN-X has recently been completed and is currently undergoing commissioning.

The new machine, ENGIN-X, (Edwards et al (2)), was designed, (Johnson et al. (3)), with one overriding principle; the instrument should be able to speedily measure lattice parameters at precisely known locations to a high precision. Meeting these requirements would benefit the broad spectrum of engineering measurements undertaken at a facility such as ENGIN-X by allowing greater throughput of components and a higher density of measurements per component. In addition new and important areas of research would be opened up including:

- Studies requiring high spatial resolution of detailed stress fields.
- Studies requiring good temporal resolution, for example, measuring the changing stress field of a component subject to variable loading or heating.

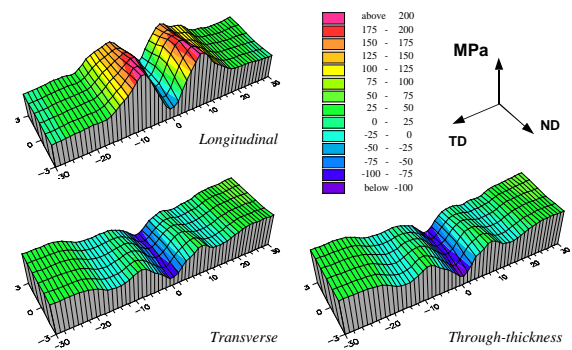

*Figure 1. Detailed maps of the residual stresses across a 7050 Aluminium Alloy MIG welded component.*

These studies in particular require both short measurement times, and the ability to place small gauge volumes with high positional accuracy. For example: the detailed 3D residual stress maps shown in Figure.1 involved making hundreds of measurements to a strain accuracy of 50µ and a spatial accuracy of 10 µm. This took over 7 days on ENGIN.

ENGIN-X offers an order of magnitude improvement in performance over ENGIN (measured using the figure of merit FOM=1/T where T is the time taken to measure a lattice parameter to a given accuracy). That the instrument hardware is capable of this performance is a necessary rather than a sufficient condition for actually achieving it in day to day operation. The routine tasks of planning, setting up, executing and analysing a measurement run need to be similarly speedy and accurate or little of the potential benefit will be realised. There is after all, little point in decreasing the measurement time from 30 minutes on ENGIN to 3 min on ENGIN-X if the setup time remains several hours on each, or designing a machine that can determine the lattice parameter to within 50 μ if there is an uncertainty of ±100μm as to the position of the measurement within the component.

Designing and implementing the computational methods to achieve speed and accuracy in these routine operations is the primary objective of the ENGIN-X software project.

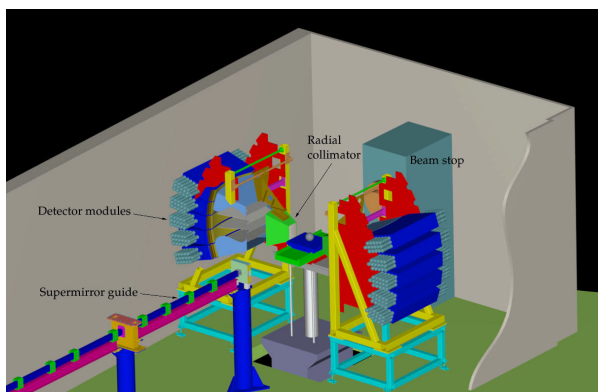

*Figure 2. A schematic of the ENGIN-X instrument.*

## 2. Analysis of a typical strain scanning measurement.

The routine operations associated with a typical stress measurement on ENGIN are:

1. *Production of a measurement plan.* The points within the component and orientations at which the stress is to be determined are identified. The exposure times needed to measure the lattice parameter to the required accuracy are estimated based on any past experience. The sequence of positioner movements required to execute the measurement run is determined manually and written as a script.
2. *Executing the measurement plan.* The script is passed to the machine control computer and each movement executed in turn. Exposure times are adjusted in accordance with a running error analysis based on the integrated neutron count for each measurement. No data analysis is possible until the run has completed.
3. *Data analysis*. Once the run has completed the data is retrieved and data reduction performed using a command line interface to the GSAS (Larson et al., (4)) Rietveld analysis package.
4. *Data Visualisation.* Data visualisation is limited to a variety of two dimensional plots in data space performed offline after transferring the data to a personal computer.

Little automation currently exists on ENGIN to perform these tasks. As a result much of this routine work needs to be performed by the instrument scientist for all but the most experienced users, this is very time-consuming in terms of manpower and prevents an optimal utilization of neutron beam time. The new software support should be such that after some initial training a visiting industrialist or researcher will be able to complete the majority of an experiment on his/her own.

The analysis below shows where this support is most needed if this aim is to be realised.

### 2.1 Computer aided measurement plans.

Defining the size and location of the measurement points, component orientations and exposure times requires considerable knowledge and experience of the material in question and of the strain scanning process. These tasks need to be performed or guided by the instrument scientist, however there remains considerable scope for software support, particularly through the provision of a convenient means of specifying arrays of measurement points within the sample, and through the automatic estimation of measurement times. The calculation of measurement times, particularly for measurements on samples of complex geometry, or for near surface measurements will help when estimating the number of points which may be measured within the limited time available for the experiment.

The calculation of machine control scripts is a task more appropriately assigned to the computer. Although theoretically simple it is fiddly, time consuming and prone to potentially expensive mistakes. Conveniently the specification of precise measurement points within the sample as discussed above renders the calculation of the control scripts a relatively trivial operation once the component has been located within the instrument.

The culmination of the planning phase then should be an automatically produced machine control script derived from accurately and conveniently defined measurement points. The script should include estimates of measurement times based on calculated neutron path lengths through the material.

Automated script production will require accurate knowledge of the position of the component in relation to the instrument; the software will need to support a method for determining this that is simple, reliable and accurate.

The intention of the support described is to provide a '*Computer aided measurement plan'*, the provision of which is a key requirement of ENGIN-X software.

### 2.2 Execution/ modification of the measurement plan.

Having produced a complete measurement plan and machine control script the task of executing the measurement run is in principle reduced to that of passing the script to the control computer. However this implies a perfect measurement plan, which in turn implies perfect knowledge of the sample; in which case there is perhaps no need to take the measurements at all!

A more realistic assessment is that most measurement plans would benefit from modification in the light of information about the stresses within the sample, if such information were available during the run. Previously, on ENGIN, it was not possible to perform any real time data analysis, hence information which might have indicated modification is not available until the plan has been completed and analysed. There is great scope for increasing efficiency by supporting real time analysis of the data as it accumulates and using this information to manually, or possible automatically, modify the measurement plan. Providing support for this feedback is another key aim of the ENGIN-X software support.

*2.3 Data analysis.*

The basic data reduction in Time of flight (TOF) neutron strain scanning consists of the time focussing of the diffraction bank (Jorgensen et al (9)), followed by a least-squares Rietveld refinement on the focussed spectrum aimed to define the lattice parameter(s) and associated error(s). At a later stage, the strain at each point of the specimen is obtained by comparing the value of the lattice parameter with that of the unstressed condition. Finally, stresses are calculated from strains using the generalized Hooke's law.

The two priorities of the software support here are *(i)* to provide a user friendly interface to the data reduction routines and *(ii)* to speed up as much as possible the process of defining the lattice parameter from the raw data. The intention is that the software should be able to perform close to real time monitoring of the evolution of the lattice parameter and its error thereby allowing tailoring of the measuring time to the desired strain accuracy (typically 50μ ). Feeding back this information to modify the temporal aspects of the ongoing measurement is one aspect of the measurement plan modification described in the previous section.

*2.4 Data visualisation.*

In a triaxial stress measurement the lattice parameters measured for a given position at different orientations are combined with the elastic constants of the material to calculate the local stress tensor. The whole process of combining different runs, calculating stresses and plotting them onto a model of the specimen was previously very time-consuming on ENGIN, preventing the inspection of the results during the experiment. Although the experienced user may still want to use his/her own plotting and analysis routines, the new software should also provide a GUI interface to simple and powerful data visualisation options; for example, displaying the data within a CAD model of the component. Ideally, the software should be able to produce strain and stress maps in real time hence producing feedback for the optimisation of the spatial aspects of the measurement plan, for example by indicating regions where additional measurements may be needed to resolve the strain field. This is the second aspect of the measurement plan modification requirement.

The software requirements described above may be summarised as providing

i. Computer Aided Measurement Plans.
ii. Automated sample alignment and orientation.
iii. Automated instrument operation.
iv. Real time primary data analysis.
v. Optimisation of data acquisition times.
vi. Automated 'first cut' strain analysis.
vii. Results visualisation.
viii. The package should be accessible to visiting industrialists and researchers after minimal training.

The figure below illustrates the relation of the new software to the previous support.

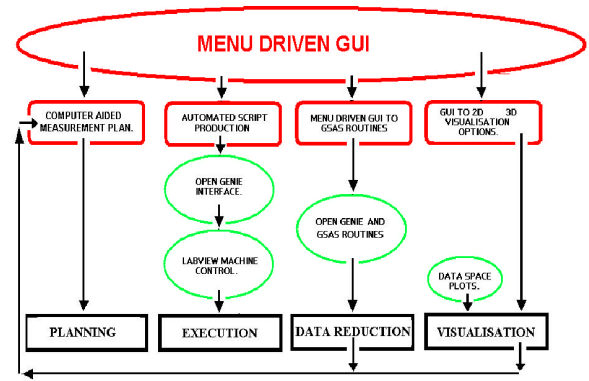

*Figure3. A schematic illustrating the scope of the new software(red) and its relation to previous support(green).*

**3. The software solution; Strain Scanning Simulation Software, (SScanSS).**

From earliest consideration it was apparent that the basis of the software should be three dimensional visualisation and simulation ie, a virtual laboratory or instrument. Modern graphics packages incorporate powerful methods for depicting and manipulating accurate representations of three dimensional objects. Such objects are constructed from polygons whose vertices are a series of surface mesh points. With this representation operations such as slicing, rotating and translating which are generally impossible for arbitrary, complex shapes become relatively simple exercises in trigonometry. A further important advantage of setting the problem in a virtual lab is that it makes maximum use of the users intuition and sense of familiarity.

The language chosen for the software was Research Systems Interactive Data Language (IDL) [ref 5]. IDL supports 3D visualisation as described as well as the writing of GUI's, data visualisation and general numeric programming all of which are required by the project. Further advantages to choosing IDL for this project are that it can talk to the ISIS in-house data manipulation and interfacing language (OpenGenie) [ref 6], plus there is expertise in IDL at ISIS. There is also some use of IDL at other neutron strain scanning facilities opening the possibility of sharing code and development effort.

One possible compromise in using IDL is that it may not be the optimum solution for complex 3D visualisation, such as that required for tomography experiments. However it is anticipated that the number of points required to adequately describe most components will be such that the performance of IDL will be sufficient.

The following presents a brief description, including illustrations, of some of the main features of the SScanSS software as it would be used to plan, execute and analyse a measurement run on an engineering component; in this case an artificial hip joint.

*3.1 The Computer aided measurement plan.*

The first steps in this section are concerned with setting up the software model of the component or sample, including fiducial and measurement points.

*3.1.1 Setting up the Virtual Component.*

The model of the component is defined by a set of surface points and connectivities. Such a model may be generated by meshing a set of surface points obtained from a coordinate measurement machine (CMM). It is hoped that the option to generate surface models via a CMM will be available on site at ISIS. Alternatively, user supplied CAD models will be used, or if neither of these options is available, the component or the relevant portion of the component will be constructed from primitives.

*3.1.2 Defining fiducial points.*

Once the model component has been defined the coordinates of a minimum of three fiducial points are specified. These fiducial points will be used to locate the model within the virtual lab.

*3.1.3 Specifying the component material.*

The next step is to specify the composition of the specimen. Polycrystalline systems with up to three component phases may be specified. The phases are selected from a library of typical engineering material files containing the crystallographic, mechanical and neutronic information necessary for Rietveld refinements and stress calculations. The output from this operation is a GSAS experiment file (with additional mechanical properties) which contains all the information required for the data reduction operations. Experienced users can choose to use their own GSAS files for defining the properties of the material. At this point in the program there is also an option to produce a simulated spectrum for comparison with the observed data spectra.

*3.1.4 Defining measurement points.*

The software offers a number of options for defining measurement points, including simple 'point and click' operations with the mouse on the screen. Figure 4 shows a two dimensional arrays of points positioned using the mouse on a section through the model. Such sections may be defined algebraically or interactively on screen.

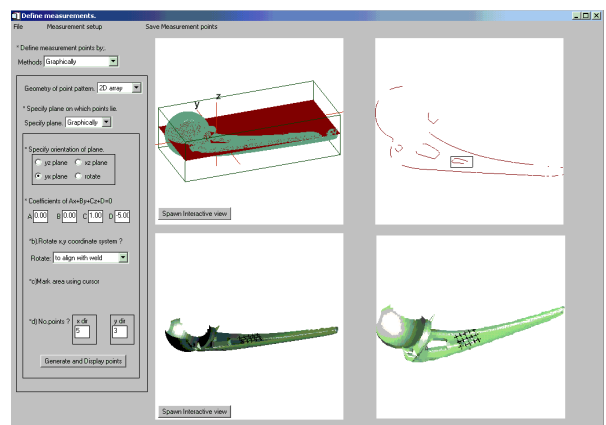

*Figure 4. A rectangular array of measurement points is defined on a plane bisecting a CMM generated mode of a artificial hip joint. The array of points is positioned using the cursor to define a rectangle within the section defined by the plane (top right).*

The set of surface points, plus fiducial and measurement points form the completed virtual component. The next step is to define the virtual instrument and to position the virtual component within it.

*3.1.5 Defining the Virtual Instrument.*

The virtual instrument consists of a collection of three dimensional objects defined in the same way as the component model. These objects represent those elements of the instrument hardware that might interact with the component; typically the incident slit, positioner and radial collimators which define the gauge volume along the incident beam direction. Other hardware, for example stress rigs or furnaces, may be included as required.

*3.1.6 Determining the location of the component within the virtual instrument.*

As the intention is to use the virtual lab to calculate the machine control scripts it is vital that the position of the model component corresponds exactly to position of the real component in the real lab. This is achieved by measuring the position of the fiducial points in the real lab and then by finding the coordinate transformation that maps the virtual fiducial points to the same positions in the virtual lab. It is intended that the first step in this operation; finding the coordinates of the real fiducial points will be achieved using triangulation with engineering theodolites, although other positioning methodologies can also be employed. A tool to support this potentially awkward operation has been written and is described in section 3.5.1 below. Figure 5. illustrates the window that supports the positioning process.

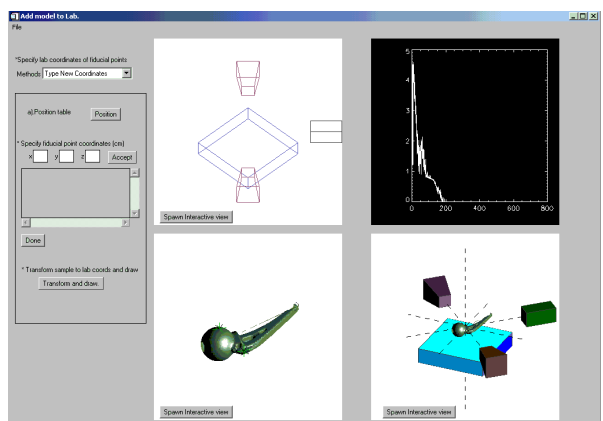

*Figure 5. The lab frame coordinates of the fiducial points are measured using theodolites and the transformation required to map the virtual component to these points is calculated using an iterative algorithm to minimise the chi squared error between the calculated and measured positions of the points. Provided the algorithm has converged to an acceptable tolerance (top right) co-location of the component and the instrument is accepted, (lower right).*

*3.1.7 Calculating the machine control script.*

Once the model component is positioned correctly in relation to the instrument the movements required to bring each of the measurement points to the gauge volume are calculated and written as a machine control script. It is also possible at this point to check that none of these movements results in a collision. This is achieved by searching for instances where lines connecting the surface points of the model component intersect the planes that define the surface of the instrument components (eg collimators) and vice versa. Using similar techniques it is also possible to calculate the neutron path length through the material, this distance may then be combined with knowledge of the scattering cross section of the material, the beam geometry and intensity to provide an estimate of the required measurement times.

*3.2 Executing the run.*

The machine control script is now passed via OpenGenie (Campbell, 2002 (7)) to the ISIS control software (Akeroyd, 2002 (8)) running on the dedicated data acquisition computer. The scan simulation is presented as illustrated in figure 6.

*3.3 Data analysis.*

On ENGIN the data processing required to calculate the value of the lattice parameter from the raw data was via a command line interface to OpenGenie and **GSAS** routines. These routines are proven, reliable and efficient and the priority for the ENGIN-X software was to provide a user friendly menu driven interface to these same routines. The new GUI gives access to OpenGenie routines which focus and normalize the detection banks, and make calls to GSAS to perform the Rietveld refinements. After one refinement has converged, the GUI provides on-line and automatic refinement of multiple runs at the click of a button. Additionally, a GUI to typical GSAS options used for strain analysis is provided in order to help with finding a converging solution. This part of the code is still under development.

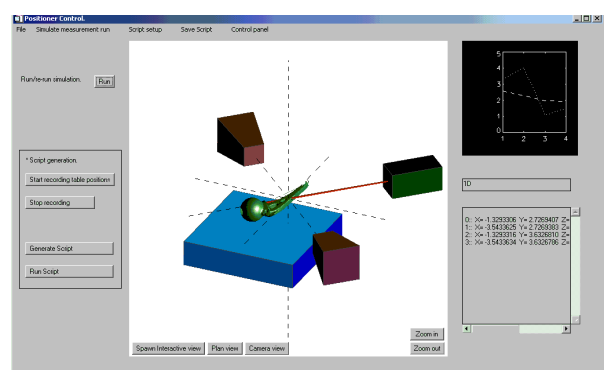

*Figure 6 The measurement run is simulated and a machine control script produced automatically (lower right). A collision check is performed for each measurement. Neutron path lengths are calculated for each detector (upper right) as a precursor to calculating exposure times.*

*3.4 Data Visualisation.*

Data visualisation options will continue to be expanded, currently simple 2D and 3D representations are available as shown in figure 7. Consultation is taking place between this project and another EPSRC funded project 'The Engineering Body Scanner' at the Manchester Materials Science Centre to ensure the use of flexible model and data formats which will enable both models and data from different centres to be exchanged or combined.

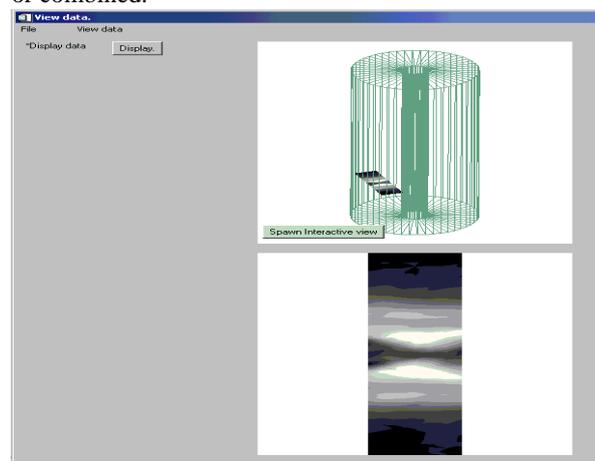

*Figure 7. Simple 2D (lower) and 3D 'in-situ' (upper) data visualisation options. All 3D object visualisations are interactive allowing rotation, zooming etc.*

*3.5 Ancillary tools.*

The description above gave a brief outline of the core components of the SScanSS package. Other tools have been written to support particularly difficult or time consuming tasks. Two of these, theodolite control and instrument calibration may be of some general interest and are described briefly below.

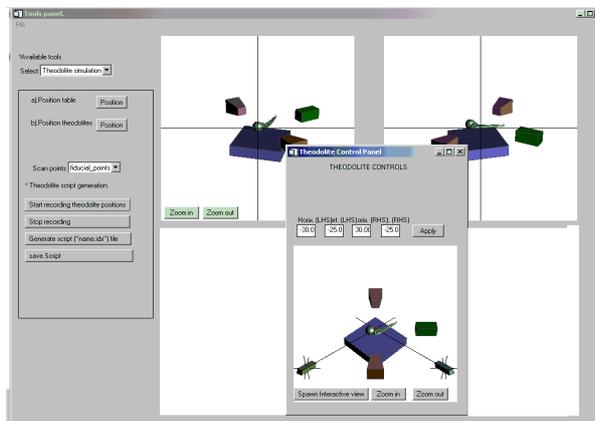

*Figure 8. The theodolite tool consists of a control panel (centre) and two simulated views, (top left and top right).*

*3.5.1 Theodolite simulation.*

The theodolite tool simulates the operation of the theodolites in the lab. For given angles of acsension and declination the view through each theodolite is anticipated. Simultaneously a triangulation is performed to determine the coordinates of the points corresponding to the theodolite cross hairs. The tool may be used to assist with the positioning task described above but also as a means of checking the position of the instrument hardware to ensure against misalignment. Theodolite control scripts may be written automatically to bring the theodolites to focus at any point that may be defined within the virtual lab.

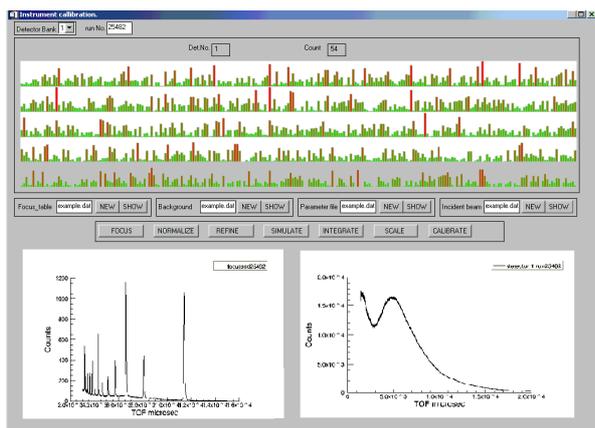

*Figure 9. Monitoring the performance of individual detector elements. Top panel shows integrated counts in each detector element. Bottom panels show individual detector recorded intensity vs time-of-flight for; a diffraction detector ,(bottom left) and the incident beam monitor (bottom right).*

*3.5.2 Machine Operation*

The software includes a number of steps to allow monitoring of the performance of the instrument, and options for the manipulation and normalisation of the raw data prior to the refinement described in 3.3. Figure 9 shows a diffraction spectrum from an individual detector element (bottom left) and the incident wavelength distribution observed in a low efficiency monitor (bottom right). The top of Figure 9 shows the integrated intensity seen in each detector element. Such a display allows monitoring of detector efficiency and performance. Detector elements can be toggled on/off using this graphical display. Other checks which will be implemented include monitoring for electronic 'spikes'.

**4. Conclusions and further work.**

We are in the process of developing a comprehensive, graphically driven suite of software to simplify the setup, running and subsequent analysis of diffraction data on the new ENGIN-X strain scanning diffractometer at ISIS.

The code is modular and acts as an interface between the acquisition control software, data analysis routines and the user in a simple manner designed to aid new academic and industrial users in conducting their experiments.

**Acknowledgements**

The authors would like to thank The Engineering Body Scanner project of The Manchester Materials Science Centre, Manchester University, for provision of the CMM scan of a hip joint.